# National research assessment exercises: a measure of the distortion of performance rankings when labor input is treated as uniform[1]

*Giovanni Abramo*[a,b,*]*, Ciriaco Andrea D'Angelo*[a] *and Marco Solazzi*[a]

[a] Laboratory for Studies of Research and Technology Transfer
School of Engineering, Dept of Management
University of Rome "Tor Vergata"

[b] National Research Council of Italy

**Abstract**

Measuring the efficiency of scientific research activity presents critical methodological aspects, many of which have not been sufficiently studied. Although many studies have assessed the relation between quality and research productivity and academic rank, not much is known about the extent of distortion in national university performance rankings when academic rank and the other labor factors are not considered as a factor of normalization. This work presents a comparative analysis that aims to quantify the sensitivity of bibliometric rankings to the choice of input, with input considered as only the number of researchers on staff, or alternatively where their cost is also considered. The field of observation consists of all 69 Italian universities active in the hard sciences. Performance measures are based on the 81,000 publications produced during the 2004-2006 triennium by all 34,000 research staff, with analysis carried out at the level of individual disciplines, 187 in total. The effect of the switch from labor to cost seems to be minimal except for a few outliers.

## 1. Introduction

In recent years, various nations have placed increasing emphasis on evaluating the production efficiency of research activity in universities and public research organizations. This has created a need for improved methods of research evaluation. Over the course of the years there was "a convergence of methods towards peer informed, metrics based, departmental level evaluation" (Hicks, 2009). The peer review approach remains central, and within this, bibliometric analysis provides useful support for the assembled panels of experts (van Raan, 2005; Rinia et al., 1998).

One of the advantages of the bibliometric approach, which is readily applicable to the hard sciences[2], is the possibility to measure labor productivity, which is a fundamental indicator of research efficiency. This factor is not measurable through peer review, in which costs and times limit the evaluation to a partial segment of the entire scientific production. As a consequence, peer-review approaches can assess only the quality of the research output submitted to evaluation. Although a certain level of correlation between output quality and productivity has been demonstrated (Abramo et al., 2009a), direct measures of productivity would permit much more than a rough approximation.

Bibliometric measurement of productivity presents two main obstacles, though. The first being the reconciliation of the different ways the authors affiliated to the same organization report it in the address. The second being the unequivocal association of publications with their true authors. It is not surprising that the literature offers few analyses, all of which are limited to a restricted number of scientific disciplines and research institutions (Macri and Dipendra, 2006; Kalaitzidakis et al., 2003; Pomfret and Wang, 2003). To the best of our knowledge, only Abramo et al. (2008a) have achieved comparative bibliometric measures of research productivity for all the hard sciences at all the universities of a national system.

Other obstacles to equitable comparison of research productivity involve the factors of production: comparison of labor productivity among various research units should be conducted at parity of other production factors and economic rents. But the factors of production, with the exception of labor, do not always permit ready measurement and accurate attribution to individual production units. It is even difficult to measure the labor factor, in hours, since the time that scientists dedicate to research varies within single universities, among institutions, and certainly between those employed at universities and those in public research institutes. It is also difficult to measure capital and certain factors that go beyond merit (such as geographic location[3], or the accumulated experience and knowledge of the scientists belonging to an institution), and to assign measurements to individual research units, with subsequent normalization, even though these factors impact directly and indirectly on output of research activity. In fact, the quality of scientific production, as measured by national peer review assessment exercises, would be influenced by these same variables[4].

---

[2] For the hard sciences, unlike the social sciences, arts and humanities, articles in international journals provide a good proxy of overall research output.
[3] Through a geographic proximity effect, concentration of public and private research organizations in a specific area can favor scientific collaboration and research productivity (Abramo et al. 2009b).
[4] Abramo et al. (2009b) demonstrate that publications in co-authorship with other organizations have a higher mean quality than those authored within a single institution. Since location affects opportunities for collaboration with other organizations it can thus have an effect on quality of output.

Limiting our attention to the relatively "more measurable" labor input, we must still consider that research staffs are composed of different academic ranks, which receive different salaries. Various scholars have examined the relationship between scientific productivity and academic rank and their studies show a significant differential in productivity with variation in rank (Prpic, 1996; Zainab, 1999; Bordons et al., 2003). As early as 1978, Blackburn et al., in a study sample of American academics, showed that full professors publish at a higher average rate than associate professors. Dickson (1983) and Kyvik (1990) have captured the same effect in their respective studies of Canadian and Norwegian universities. There has been less study of the relationship between quality of output and academic rank. Bordon et al. (2003) analyze the impact of publications by Spanish Research Council scientists by gender and professional category, in two specific areas: Natural resources and Chemistry. They show that the average impact factor of journals in which full professors publish their articles is higher than that for publications by the lower academic ranks. Abramo et al. (2009c) extend the analysis to all the hard sciences and demonstrate that Italian full professors average more publications than associate professors (and these more than assistant professors), and also in journals with a higher impact factor. A further study by Abramo et al. (2009d) demonstrates a strong correlation between productivity and impact, meaning that the scientific production by the most productive scientists is also, on average, of greater quality. Ben-David (2009) showed that Israeli economists with the rank of professor receive on average more citations than their colleagues with lower ranks. These studies confirm the expectation that quality of output reflects academic rank.

A consequence is that university rankings based on productivity or on quality by uniform labor unit will clearly favor organizational units with a greater concentration of higher roles. If national research assessment exercises do not take this effect into account, leaving resulting distortions in their rankings, there could be possible dangerous effects on allocation of public funds and on the image of the institutions observed. This is the case for the example of the first and only Italian national research evaluation exercise, VTR, and for the subsequent allocation of the portion of public financing that is partially based on VTR rankings. These rankings have not accounted for the varying presence of staff ranks among different universities.

The present study intends to measure the extent of distortions in national performance rankings of research institutions when academic rank, and relevant salaries, are not taken into account. We do not expect that such distortions are very high on average, because of two main reasons. The first being that the concentrations of academic ranks are similar across universities, with few possible exceptions especially among younger universities. The second being that academic salaries in Italy are fixed at national level and depend only on role and seniority, not on merit.

Using bibliometric techniques, we compare two different rankings of research productivity in Italian universities: one which considers the labor factor as homogenous and one which considers the differing academic rank of the research staff. We carry out such comparisons at two different levels: at a detailed scientific sectorial level; and at more aggregated discipline level. In each discipline and scientific sector within the discipline, we measure the changes in the above said rankings, and provide the relevant statistics. Considering that data about the academic ranks and salary ranges of the Italian university personnel are available, and also that the proportions of such personnel in organizational units, although similar, are not the same, we propose that in such a case, the comparison of research productivity by "unit of cost" would be more equitable than comparison by unit of labor, all other limitations of productivity measurements remaining the same.

The following section of this paper describes the field of observation for the study, the dataset and the methodology used. Section 3 presents the results of the analysis. The final section offers a discussion of the results and the authors' concluding considerations.

## 2. Methodological approach

Research activity is an input-output production process in which the inputs consist of human and financial resources, scientific instruments, materials, etc., and where outputs have a complex character of both tangible nature (publications, patents, conference presentations, etc.) and intangible nature (personal knowledge, consulting activity, etc.). The knowledge production function has a multi-input and multi-output character. This in turn creates a multi-faceted problem when it comes to measuring the scientific productivity of labor, and requires scholars to make precise choices in methodology.

In this work, measuring the scientific productivity of Italian universities in the hard sciences, we first consider input only as the number of researchers involved, but subsequently also consider their relative cost.

Concerning output, there are multiple forms of codification for new knowledge produced by research activity. Having limited the field of analysis to the hard sciences, we choose scientific publications as a proxy for research output, which certainly finds support in the literature (Moed et al., 2005). The research productivity of individual scientists is not normalized to their actual hours of research time or to other productive factors, since there is a complete lack of data that can be attributed to the level of individuals.

### 2.1 Dataset

The data used in the study are obtained from the Observatory on Public Research in Italy (ORP), a bibliometric database maintained by the authors and derived from Thomson Reuters' Web of Science (WoS). The ORP provides a census of WoS indexed scientific production since 2001, from all research institutions situated in Italy. Beginning from the ORP data, this study extracted all publications (articles and reviews) authored by researchers in Italian universities for the period 2004-2006. A reconciliation of the different denominations of the same universities followed[5]. Finally, using a complex algorithm for disambiguation of the precise identity of the authors, each publication was attributed to the university scientists who wrote it[6].

In the Italian university system, each researcher is assigned to a single official scientific disciplinary sector (SDS). For the hard sciences, there are 183 SDSs[7], grouped into 8 disciplinary areas (UDAs): Mathematics and computer sciences; Physics; Chemistry; Earth sciences; Biology; Medicine; Agricultural and veterinary sciences; and Industrial and information engineering[8]. The census by author name permits attribution of measures of output to individual researchers, and then by aggregation to the SDS and UDA of a university. The methods used overcome considerable obstacles and provide levels of accuracy that have not previously been attained in large-scale studies in the literature. When one observes large populations of scientists, the number of homonyms among their names is very high (in the Italian academic system 12% of the 60,000 scientists have names that are homonyms), and the task of their disambiguation within acceptable margins of error is formidable. This is why bibliometrics-based studies have generally been carried out at aggregated levels of analysis, such as at the level of entire universities. When they are conducted at the levels of single scientists or research group they are limited to one or few organizations or scientific disciplines, in which case it is possible to disambiguate manually. Disambiguation can not be done manually in the case of an evaluation of an entire national research system, where an enormous quantity of data is involved. However this step is required in order to avoid distortions in productivity measurement caused by several factors: i) the differing distribution of resources among the various scientific areas of each university, ii) varying degrees of publication and citation "fertility" among scientific disciplines; iii) variation in the data source in terms of its differing coverage of the range of journals published in each disciplinary area; and iv) the researchers generally publishing in more than one subject category.

---

[5] On the subject of address reconciliation, Geuna and Martin (2003) report: "… The main problem consists in having to 'clean up' institutional addresses, a task that can take many person-years of effort".
[6] At this time, for disambiguation of authorship of the 215,000 Italian academic publications indexed in the WoS between 2001 and 2007, the harmonic average of precision and recall (F-measure) is close to 95% (2% sampling error, 98% confidence interval). Further details are reported in Abramo et al. (2008a).

For the 2004-2006 triennium, this study concerns the 69 Italian universities active in the 183 hard science SDSs, representing a total of 34,000 research staff with over 81,000 publications. The official database of the Ministry of Education, Universities and Research (MIUR)[9] was used to provide a census of all university research personnel and their roles. This ministry is responsible for the recognition of university status, allocation of regular operating funding, and the control and evaluation of university function.

Data concerning salary costs for research personnel were obtained from the DALIA[10] database, which is also maintained by the MIUR. The current Italian university system provides that research personnel are assigned to three roles: full professors, associate professors and assistant professors. Definitive confirmation of an individual's rank arrives after a three year "probationary" appointment, following an examination of the individual's performance. The university system also includes a small number of "research assistants", a role which is being eliminated, and which resembles that of an assistant professor. Table 1 shows the numbers, total costs and average cost per rank of these personnel, for the triennium.

[Table 1]

Full professors compose 29.5% of university personnel but represent 40.6% of total personnel costs. Assistant professors compose the largest portion of personnel, at 37.7%, but represent only 27.4% of the entire cost. The last column of Table 1 presents the figures for average costs per academic rank, which are used in the subsequent elaborations of productivity on the basis of cost.

---

[7] The complete list is available at http://www.miur.it/atti/2000/alladm001004_01.htm
[8] "Civil engineering and architecture" UDA was not considered because the WoS does not cover the full range of research output in this area.
[9] http://cercauniversita.cineca.it/php5/docenti/cerca.php
[10] https://dalia.cineca.it/php4/inizio_access_cnvsu.php

**2.2 Indicators**

For each publication in the dataset, the study considers an indicator of quality defined as Article Impact Index, measured on a 0–100 percentile scale, according to the citation[11] distribution for publications of the same type and year falling in the same ISI subject category[12]. A value of 90 indicates that 90% of the articles (or reviews) of the same year, falling in the same ISI category, have a lower number of citations than the article (or review) considered. In this way the quality measurement distortions due to the different citation fertilities among subject categories are limited.

The indicator for evaluation of the bibliometric output of the researchers in the various university SDSs is Fractional Scientific Strength. This is given by the sum of the publications achieved by the researchers of a single university SDS, with each publication weighted according to its Article Impact Index and normalized according to the number of organizations to which the coauthors belong. With this method it is possible to consider all dimensions relevant to output: the quantitative (through number of publications), qualitative (through Article Impact Index) and the dimension of contribution (through the count of co-authorship).

The productivity of a particular university SDS is given by the ratio of Fractional Scientific Strength to the input factor for the same SDS. For the productivity per labor unit (LP), the input factor considered is simply the number of scientists present in the SDS, while for the calculation of productivity per unit of cost (CP) the input factor considered is the overall cost of research staff at the SDS, derived from the parameters indicated in the last column of Table 1.

Continuing on from the level of the SDS, the productivity values for a full university UDA are then obtained by aggregation, after standardization and weighting. Productivity measures of each university in each SDS are therefore standardized to the national mean in the same SDS. This standardization serves to eliminate bias due to the different publication and citation rates of the SDSs within a single UDA. Data weighting instead takes account of the variation in representativity, in terms of personnel numbers and costs, of the SDS represented within each UDA (Abramo et al., 2008b). For a generic university we thus have:

$$LP_j = \sum_{s=1}^{n_j} \left( \frac{LP_s}{\overline{LP_s}} \cdot \frac{Add_s}{Add_j} \right)$$

where:
  $LP_j$ = productivity per labor unit in UDA $j$,
  $LP_s$ = productivity per labor unit in SDS $s$,

---

[11] The basic assumption of bibliometrics, e.g. the level of citation which corresponds to a quantum of research quality, has been criticized by few scholars (Warner, 2000). In this study though we are not interested in absolute ratings, but in switch of rankings when passing form labor input to cost input.
[12] The ISI subject categories are the scientific disciplines that the WoS uses for classification of articles.

$\overline{LP_s}$ = national mean of productivity per labor unit in SDS *s*,
$Add_s$ = number of scientists in the university considered in SDS *s*,
$Add_j$ = number of scientists in the university considered in UDA *j*,
$n_j$ = number of SDSs in the university considered in UDA *j*.

Analogously:

$$CP_j = \sum_{s=1}^{n_j}\left(\frac{CP_s}{\overline{CP_s}} \cdot \frac{Add_s}{Add_j}\right)$$

where:
$CP_j$ = productivity per unit of cost in UDA *j*,
$CP_s$ = productivity per unit of cost in SDS *s*,
$\overline{CP_s}$ = national mean of productivity per unit of cost in SDS *s*.

## 3. Results

As described above, ratings of productivity for Italian universities were calculated per labor unit and unit of cost, for the 2004-2006 triennium, and then used to obtain rankings. In the following, changes in rankings when switching from measure of productivity per labor unit to unit of cost, are shown at the UDA and SDS levels. Table 2 presents the variations under the two methods of ranking, as recorded for each UDA in each university[13].

[Table 2]

Table 3 presents further statistics concerning the distribution of the rankings under the two different methods, by UDA, for the field of observation. As expected, it is readily apparent that there is a very high correlation between the two rankings, in all areas (last column of Table 3). The coefficient of correlation varies from a minimum of 0.972 for Biology to a maximum of 0.996 for Agricultural and veterinary sciences. But at the same time, the variations in ranking between the two methods are also quite substantial: the number of universities for which the ranking changes under the two methods ranges from a maximum of 86.4% for Physics to a minimum of 36.5% for Agricultural and veterinary sciences. This last UDA shows the strongest correlation between the two rankings: 33 of the 52 universities maintain a constant ranking under the two methods. It also presents the lowest values for the other statistics presented in Table 3: the greatest shift in position is only 3 places, seen at 3 distinct universities (Sassari, Teramo and Udine) while the average shift in rank is less than one (0.615) and the median is zero. The maximum mean value of change in ranking is seen in the biology UDA (2.667), followed by industrial and information engineering (2.258), physics (2.237) and chemistry (2.207). The chemistry UDA offers the extreme case of a university that shifts 17 positions under the two methods of ranking. Other wide jumps in ranking occur in Physics, where the University of Reggio Calabria "Mediterranean" gains 15 places under the CP classification, with respect to its ranking for LP. In Industrial and information engineering there is a shift of the same magnitude: in this case the University of Rome "Foro Italico" loses 15 positions under the CP classification compared to its LP ranking. In Biology, the maximum variation in ranking is 13 positions, and concerns three universities: The University of Teramo gains positions, while the universities of Milan "Vita-Salute San Raffaele" and Venice "Ca' Foscari" lose the same number. The same extent of shift occurs in Earth sciences, for the University of Trent, which loses 13 positions when classified for CP as compared to LP.

---
[13] The rankings of the Italian peer-review VTR were carried out at the UDA level.

[Table 3]

Table 4 presents data on the calculation and ranking of productivity for universities active in the Chemistry UDA, as an in-depth example of one of the areas that presents greater shifts in rankings under the two methods. In this UDA, 42 of the 58 total universities show a different ranking under the classification by LP and by CP. Of these 42, 39 show variations in ranking with absolute values less than or equal to 4. The maximum shift is 17 positions, as noted above, for the University of Teramo: this university, a rather young one, jumps from 40th position under LP to 24th under CP. The staff complement here consists of 4 scientists (averaged over the triennium) with an average cost of €64,400, which is the least among all the universities active in the UDA, since there are no full professors present. The situation is similar for the University of Cassino, which places in 40th position under LP but rises to 24th position under CP. The trend is the opposite for the University of Catania, with a heavy concentration of top-ranked personnel among its 107 scientists (mean cost per scientist: €94,700), which contributes to losing 7 positions under the classification by CP compared to that for LP. Only the International School for Advanced Studies of Trieste shows a higher value of mean cost per scientist, at €98,900. In general, there is a significant correlation (-0.739) between the variation in LP and CP ranking and the mean cost per member of research staff in each university, active in this UDA.

At the SDS level, Table 5 presents data on the calculation and ranking of productivity for the 45 universities active in the Pharmacology SDS of the Biology UDA, as an example of the variation that may be observed at a more detailed level. The shifts in ranking seem less than at the level of UDA: the mean value of shift is 1.33, in absolute value, with a median of 1. The maximum variation is seen for the University of Milan "Vita-Salute San Raffaele" which drops from fifth position for LP to 13$^{th}$ for CP. The maximum "positive" shift in direction is seen for the Second University of Naples, which gains 4 positions, moving from 38$^{th}$ ranked for LP to 34$^{th}$ ranked for CP. In total, eight universities show increases in ranking that are equal to or greater than 3 places, for CP, while 12 universities do not show any change in position.

[Table 4]
[Table 5]

4. Conclusions

Bibliometric techniques permit the measurement of research productivity of universities and public research institutions. Comparative measures of labor productivity should be conducted under parity of other factors of production, but these factors are difficult to measure and attribute to individual scientists. The first and only Italian national exercise of research evaluation, based on peer review techniques, treated the labor factor as uniform, meaning that the comparative quality of organizational units was not normalized to take account of variations in distribution of academic rank. This may occur again and in other countries as well. The current study illustrates the number and extent of distortions which occur when the labor factor is treated as uniform in the Italian university system. Other literature on the argument indicates that there is a significant difference in average productivity among academic ranks, which, when the labor factor is considered uniform, results in more favorable evaluations for universities with greater concentration of full professors.

The proposed study compared rankings of productivity for Italian universities with respect to labor unit and unit of cost. The analysis was conducted from the bottom up, beginning with the identification of the authorship of over 81,000 publications by all university 34,000 scientists working in the hard sciences, then by aggregation at the level of the scientific disciplinary sectors in individual universities and at the further level of disciplinary area. At both these levels there is a strong correlation between the two measures of productivity, but also some variations in rankings, especially in reference to a number of outliers that show substantial shifts in rank for "cost" productivity as compared to labor productivity. This occurs for universities where the personnel complement is notably imbalanced in favor of higher or lower academic ranks, and which are therefore unavoidably favored or disfavored by the assessment methodology that does not take account of the representation of research staff by academic rank.

The measurements proposed do not take account of variations in the time dedicated to research by the staff members, although teaching load and other institutional duties are not necessarily equally divided. Nor does the methodology consider the capital available to the organizational units under observation, or other factors external to merit that could impact on quantity and quality of scientific production.

Even with these cautionary notes, the study provides a useful indication of how to proceed towards research assessments that are more robust and exhaustive than those of the current state of the art. In particular, the study proposes an improvement in measurement of labor productivity that should be useful in support systems for the decisions of those who, at various levels, are responsible for the management and evaluation of research institutions and research systems. While ranking distortions due to overlooking academic rank, result negligible on average at aggregated levels of analysis, such as discipline level, they should be more noticeable at the single scientist or research group levels. The authors intend to investigate this in the future, to the benefit of those universities that implement incentive systems based on research performance.

95-107.

van Raan A.F.J., (2005). Fatal attraction: Conceptual and methodological problems in the ranking of universities by bibliometric methods. *Scientometrics*, 62(1): 133-143.

Warner J., (2000). A critical review of the application of citation studies to the Research assessment Exercises, *Journal of Information Science*, 26(6): 453-459

Zainab A.N., (1999). Personal, academic and departmental correlates of research productivity: a review of literature. *Malaysian Journal of Library & Information Science*, 4(2): 73–110.

| Academic rank | Number | | Total cost (M€) | | Average cost (k€) |
|---|---|---|---|---|---|
| Full professors (confirmed) | 8,475 | (24.8%) | 1,054.9 | (35.3%) | 124.5 |
| Full professors (probationary) | 1,599 | (4.7%) | 158.1 | (5.3%) | 98.9 |
| *Sub-tot.* | *10,074* | *(29.5%)* | *1,213.0* | *(40.6%)* | |
| Associate professors (confirmed) | 8,497 | (24.9%) | 762.9 | (25.6%) | 89.8 |
| Associate professors (probationary) | 2,474 | (7.2%) | 172.2 | (5.8%) | 69.6 |
| *Sub-tot.* | *10,971* | *(32.1%)* | *935* | *(31.3%)* | |
| Assistant professors (confirmed) | 10,500 | (30.8%) | 711.8 | (23.8%) | 67.8 |
| Assistant professors (probationary) | 2,353 | (6.9%) | 107.0 | (3.6%) | 45.5 |
| *Sub-tot.* | *12,853* | *(37.7%)* | *819* | *(27.4%)* | |
| Research assistants (obsolete rank) | 238 | (0.7%) | 18.4 | (0.6%) | 77.2 |
| *Total* | *34,136* | - | *2,985.2* | - | - |

*Table 1: Data concerning Italian university personnel, mean values 2004-2006.*

| | UDA* | | | | | | | |
|---|---|---|---|---|---|---|---|---|
| University | 1 | 2 | 3 | 4 | 5 | 6 | 7 | 8 |
| Academic institute of Architecture in Venice | 2 | NA | NA | 0 | 0 | 0 | 0 | 0 |
| International School for Advanced Studies of Trieste | -1 | -1 | 0 | NA | 0 | NA | NA | 0 |
| Polytechnic University of Ancona | 0 | 0 | -4 | 0 | 0 | -3 | 0 | -4 |
| Polytechnic University of Bari | -1 | -1 | 0 | 6 | 0 | NA | NA | 0 |
| Polytechnic University of Milan | 0 | 1 | 1 | 0 | 0 | 0 | NA | -2 |
| Polytechnic University of Turin | -1 | 3 | 2 | 0 | 0 | NA | NA | 0 |
| Sacred Heart Catholic University | -2 | -2 | NA | NA | 6 | 8 | 2 | 0 |
| Scuola Normale Superiore in Pisa | -1 | -3 | 0 | NA | 12 | NA | NA | NA |
| Scuola Superiore St.Anna in Pisa | NA | NA | NA | NA | -10 | -3 | 0 | -1 |
| Second University of Naples | 1 | -1 | 1 | 1 | -1 | 0 | 0 | 0 |
| University "Bocconi" in Milan | 0 | NA | NA | NA | NA | NA | NA | 0 |
| University of Rome "Roma Tre" | 2 | -2 | 0 | -3 | -2 | -3 | 0 | -1 |
| University of Bari | 0 | -2 | -3 | 1 | 1 | 1 | -1 | -1 |
| University of Basilicata | 5 | -1 | 0 | 0 | -1 | NA | 0 | 1 |
| University of Benevento "Sannio" | 9 | 7 | 0 | -1 | -2 | 0 | 0 | 1 |
| University of Bergamo | 0 | -1 | 1 | NA | NA | NA | NA | 3 |
| University of Bologna | -3 | 2 | -2 | -2 | 1 | 0 | 1 | 0 |
| University of Bolzano | -1 | NA | NA | NA | NA | NA | NA | 9 |
| University of Brescia | 0 | 0 | 4 | 0 | 5 | 0 | -2 | 6 |
| University of Cagliari | -1 | 3 | 4 | 0 | 1 | 0 | 0 | 5 |
| University of Calabria | -1 | 5 | -1 | 0 | 1 | 1 | 0 | 0 |
| University of Camerino | -2 | -4 | -2 | 3 | -3 | 0 | 1 | 0 |
| University of Cassino | 9 | 1 | 16 | NA | 7 | 0 | 0 | 1 |
| University of Castellanza "Carlo Cattaneo" | NA | NA | NA | NA | NA | NA | NA | 1 |
| University of Catania | -1 | -2 | -7 | 4 | -2 | 0 | 0 | -4 |
| University of Catanzaro "Magna Grecia" | NA | -1 | 0 | NA | -1 | 1 | 0 | 7 |
| University of Chieti "Gabriele D'Annunzio" | 6 | 2 | 2 | 0 | -3 | -2 | 0 | 0 |
| University of Eastern Piedmont "A. Avogadro" | -6 | 0 | 0 | 0 | 1 | -3 | NA | NA |
| University of Ferrara | -5 | 1 | -2 | 1 | -3 | -1 | 0 | 1 |
| University of Florence | -2 | -2 | -2 | -1 | 1 | 0 | 2 | -4 |
| University of Foggia | NA | 4 | 2 | NA | 1 | 3 | 2 | NA |
| University of Genova | -3 | -2 | -1 | 0 | -5 | 1 | NA | -2 |
| University of L'Aquila | 1 | -1 | -4 | -5 | -1 | 0 | 0 | -7 |
| University of Lecce "Salento" | 2 | 1 | 2 | 3 | 3 | 0 | NA | 5 |
| University of Macerata | 0 | NA | NA | NA | 0 | -1 | 0 | NA |
| University of Messina | -1 | -1 | -3 | 0 | 3 | 0 | 0 | 0 |
| University of Milan | 2 | 2 | -2 | -1 | 1 | -1 | -1 | 0 |
| University of Milan "Bicocca" | 0 | -4 | 1 | 0 | 4 | 2 | -1 | -1 |
| University of Milan "Vita-Salute San Raffaele" | NA | NA | NA | NA | -13 | 0 | NA | NA |
| University of Modena and Reggio Emilia | 0 | -2 | -2 | 0 | -3 | -1 | 2 | 2 |
| University of Molise-Campobasso | 2 | 11 | 1 | 3 | 2 | 0 | 0 | 0 |
| University of Naples "Federico II" | -1 | -1 | -2 | 0 | 1 | 1 | 1 | -4 |
| University of Naples "L'Orientale" | NA | NA | NA | NA | NA | 0 | NA | NA |
| University of Naples "Parthenope" | 0 | 0 | 0 | -2 | 1 | 3 | 0 | -1 |

|  | UDA* | | | | | | | |
|---|---|---|---|---|---|---|---|---|
| University | 1 | 2 | 3 | 4 | 5 | 6 | 7 | 8 |
| University of Padua | -3 | -4 | -2 | 0 | 1 | -2 | 0 | -3 |
| University of Palermo | 4 | 0 | 0 | 0 | 0 | 1 | 0 | 1 |
| University of Parma | 0 | -2 | 0 | 3 | 0 | -1 | 1 | 0 |
| University of Pavia | 0 | -2 | -3 | 3 | -2 | -2 | 0 | -5 |
| University of Perugia | 3 | -3 | -2 | 1 | -3 | 2 | 0 | -1 |
| University of Pisa | -1 | -1 | -3 | 1 | 1 | 0 | -2 | -5 |
| University of Reggio Calabria "Mediterranean" | 8 | 15 | 0 | NA | 0 | NA | 0 | 5 |
| University of Rome "Campus Bio-medico" | NA | 1 | -3 | NA | 0 | 3 | NA | 1 |
| University of Rome "Foro Italico" | 0 | NA | NA | NA | 0 | 1 | NA | -15 |
| University of Rome "La Sapienza" | -1 | -2 | -2 | -2 | 0 | 4 | 0 | -2 |
| University of Rome "Maria SS.Assunta" | NA | NA | NA | NA | NA | -5 | NA | NA |
| University of Rome "Tor Vergata" | -5 | -2 | -1 | NA | 0 | 2 | 0 | -1 |
| University of Salerno | 1 | 5 | 2 | -1 | -4 | 0 | 0 | 4 |
| University of Sassari | 0 | -1 | 4 | 0 | 1 | -1 | -3 | 5 |
| University of Siena | -3 | 0 | -3 | 1 | -1 | 0 | 0 | 2 |
| University of Teramo | 0 | 1 | 17 | NA | 13 | 0 | 3 | 8 |
| University of Trent | -1 | -1 | 4 | -13 | 9 | -5 | 0 | 0 |
| University of Trieste | -2 | -7 | -3 | -2 | -2 | -1 | 0 | 0 |
| University of Turin | -3 | -2 | 0 | 1 | 0 | -1 | 1 | -1 |
| University of Udine | 0 | 0 | -1 | -3 | -2 | 0 | -3 | -2 |
| University of Urbino "Carlo Bo" | 1 | 1 | 0 | 3 | 4 | 2 | 0 | 2 |
| University of Varese "Insubria" | 1 | -3 | 0 | 1 | 3 | 0 | 0 | 0 |
| University of Venice "Ca' Foscari" | -7 | -1 | -3 | 1 | -13 | NA | 0 | -2 |
| University of Verona | 0 | -1 | 0 | NA | -1 | 0 | -1 | 0 |
| University of Viterbo "Tuscia" | 0 | 0 | -1 | -1 | -6 | 0 | -2 | -1 |

*Table 2: Variations in ranking when switching from measures of productivity per labor unit (LP) to unit of cost (CP), for Italian universities, by university disciplinary area (UDA), 2004-2006 data. "NA" means that there are no scientists in the UDA.*

*\* 1 = Mathematics and computer sciences; 2 = Physics; 3 = Chemistry; 4 = Earth sciences; 5 = Biology; 6 = Medicine; 7 = Agricultural and veterinary sciences; 8 = Industrial and information engineering*

| UDA | Variations | % | Max | Mean | Median | Std Dev. | Correlation |
|---|---|---|---|---|---|---|---|
| Mathematics and computer sciences | 43 out of 61 | 70.5 | 9 | 1.934 | 1 | 2.301 | 0.985 |
| Physics | 51 out of 59 | 86.4 | 15 | 2.237 | 2 | 2.589 | 0.980 |
| Chemistry | 42 out of 58 | 72.4 | 17 | 2.207 | 2 | 3.105 | 0.974 |
| Earth sciences | 30 out of 48 | 62.5 | 13 | 1.542 | 1 | 2.231 | 0.981 |
| Biology | 48 out of 63 | 76.2 | 13 | 2.667 | 2 | 3.379 | 0.972 |
| Medicine | 33 out of 58 | 56.9 | 8 | 1.241 | 1 | 1.604 | 0.993 |
| Agricultural and veterinary sciences | 19 out of 52 | 36.5 | 3 | 0.615 | 0 | 0.932 | 0.996 |
| Industrial and information engineering | 43 out of 62 | 69.4 | 15 | 2.258 | 1 | 2.816 | 0.979 |

*Table 3: Variations in ranking statistics when switching from measures of productivity per labor unit (LP) to unit of cost (CP), by university disciplinary area (UDA), 2004-2006 data.*

| University | Average cost (k€) | LP abs.val. | LP rank | CP abs.val. | CP rank | Rank variation |
|---|---|---|---|---|---|---|
| University of Teramo | 64.4 | 0.937 | 28 | 1.262 | 11 | 17 |
| University of Cassino | 67.8 | 0.736 | 40 | 0.961 | 24 | 16 |
| University of Brescia | 72.7 | 1.400 | 9 | 1.696 | 5 | 4 |
| University of Trent | 81.2 | 0.803 | 35 | 0.866 | 31 | 4 |
| University of Cagliari | 86.8 | 0.726 | 44 | 0.728 | 40 | 4 |
| University of Sassari | 85.9 | 0.715 | 45 | 0.727 | 41 | 4 |
| University of Salerno | 79.4 | 1.312 | 10 | 1.441 | 8 | 2 |
| Polytechnic University of Turin | 85.4 | 1.036 | 20 | 1.076 | 18 | 2 |
| University of Lecce "Salento" | 81.5 | 0.760 | 37 | 0.820 | 35 | 2 |
| University of Chieti "Gabriele D'Annunzio" | 82.7 | 0.630 | 49 | 0.662 | 47 | 2 |
| University of Foggia | 78.7 | 0.517 | 52 | 0.570 | 50 | 2 |
| University of Milan "Bicocca" | 86.5 | 1.155 | 14 | 1.171 | 13 | 1 |
| University of Molise-Campobasso | 87.4 | 0.977 | 23 | 0.982 | 22 | 1 |
| Second University of Naples | 89.3 | 0.893 | 31 | 0.872 | 30 | 1 |
| Polytechnic University of Milan | 87.9 | 0.680 | 46 | 0.682 | 45 | 1 |
| University of Bergamo | 85.8 | 0.505 | 53 | 0.520 | 52 | 1 |
| International School for Advanced Studies of Trieste | 98.9 | 4.727 | 1 | 4.191 | 1 | 0 |
| University of Benevento "Sannio" | 71.4 | 2.379 | 2 | 2.975 | 2 | 0 |
| University of Catanzaro "Magna Grecia" | 84.1 | 2.152 | 3 | 2.208 | 3 | 0 |
| University of Verona | 87.2 | 1.999 | 4 | 2.008 | 4 | 0 |
| University of Eastern Piedmont "A. Avogadro" | 85.7 | 1.225 | 12 | 1.242 | 12 | 0 |
| University of Basilicata | 88.1 | 1.146 | 16 | 1.136 | 16 | 0 |
| University of Turin | 89.7 | 1.053 | 19 | 1.027 | 19 | 0 |
| University of Parma | 89.2 | 0.881 | 32 | 0.863 | 32 | 0 |
| University of Varese "Insubria" | 85.0 | 0.742 | 38 | 0.754 | 38 | 0 |
| University of Palermo | 88.8 | 0.729 | 43 | 0.713 | 43 | 0 |
| University of Urbino "Carlo Bo" | 79.8 | 0.517 | 51 | 0.557 | 51 | 0 |
| Polytechnic University of Bari | 87.4 | 0.424 | 54 | 0.430 | 54 | 0 |
| University of Rome "Roma Tre" | 81.8 | 0.402 | 55 | 0.417 | 55 | 0 |
| University of Reggio Calabria "Meditteranean" | 92.1 | 0.358 | 56 | 0.334 | 56 | 0 |
| Scuola Normale Superiore in Pisa | 67.8 | 0.142 | 57 | 0.182 | 57 | 0 |
| University of Naples "Parthenope" | 77.8 | 0.136 | 58 | 0.155 | 58 | 0 |
| University of Calabria | 85.8 | 1.589 | 5 | 1.623 | 6 | -1 |
| University of Rome "Tor Vergata" | 84.6 | 1.478 | 6 | 1.533 | 7 | -1 |
| University of Genova | 89.6 | 0.736 | 41 | 0.715 | 42 | -1 |
| University of Viterbo "Tuscia" | 90.1 | 0.659 | 47 | 0.632 | 48 | -1 |
| University of Udine | 92.0 | 0.637 | 48 | 0.597 | 49 | -1 |
| University of Ferrara | 90.8 | 1.474 | 7 | 1.410 | 9 | -2 |
| University of Florence | 89.3 | 1.405 | 8 | 1.371 | 10 | -2 |
| University of Perugia | 93.3 | 1.218 | 13 | 1.137 | 15 | -2 |
| University of Bologna | 88.9 | 1.154 | 15 | 1.132 | 17 | -2 |
| University of Naples "Federico II" | 89.0 | 0.995 | 21 | 0.973 | 23 | -2 |
| University of Modena and Reggio Emilia | 87.8 | 0.965 | 24 | 0.955 | 26 | -2 |
| University of Milan | 89.5 | 0.963 | 25 | 0.937 | 27 | -2 |
| University of Padua | 90.3 | 0.959 | 26 | 0.926 | 28 | -2 |
| University of Rome "La Sapienza" | 91.8 | 0.938 | 27 | 0.889 | 29 | -2 |
| University of Camerino | 90.2 | 0.734 | 42 | 0.705 | 44 | -2 |
| University of Siena | 92.9 | 1.232 | 11 | 1.145 | 14 | -3 |
| University of Messina | 92.7 | 1.106 | 17 | 1.026 | 20 | -3 |
| University of Trieste | 92.1 | 1.074 | 18 | 1.018 | 21 | -3 |
| University of Pisa | 89.8 | 0.981 | 22 | 0.958 | 25 | -3 |
| University of Pavia | 91.5 | 0.841 | 33 | 0.800 | 36 | -3 |
| University of Venice "Ca' Foscari" | 90.6 | 0.815 | 34 | 0.785 | 37 | -3 |
| University of Bari | 91.7 | 0.784 | 36 | 0.744 | 39 | -3 |
| University of Rome "Campus Bio-medico" | 91.5 | 0.524 | 50 | 0.506 | 53 | -3 |
| University of L'Aquila | 94.7 | 0.931 | 29 | 0.859 | 33 | -4 |
| Polytechnic University of Ancona | 94.4 | 0.894 | 30 | 0.828 | 34 | -4 |
| University of Catania | 94.7 | 0.741 | 39 | 0.681 | 46 | -7 |

*Table 4: Comparison between productivity per labor unit (LP) and unit of cost (CP) for Italian universities, for the Chemistry UDA, 2004-2006 data.*

| University | Average cost (k€) | LP abs.val. | LP rank | CP abs.val. | CP rank | Rank variation |
|---|---|---|---|---|---|---|
| Second University of Naples | 74.6 | 26.181 | 38 | 0.351 | 34 | 4 |
| University of Varese "Insubria" | 81.2 | 59.593 | 9 | 0.734 | 6 | 3 |
| University of Urbino "Carlo Bo" | 81.3 | 57.794 | 10 | 0.711 | 7 | 3 |
| Sacred Heart Catholic University | 81.5 | 52.088 | 14 | 0.639 | 11 | 3 |
| University of Bari | 78.2 | 31.361 | 32 | 0.401 | 29 | 3 |
| University of Perugia | 83.1 | 43.667 | 19 | 0.526 | 17 | 2 |
| University of Turin | 83.4 | 41.386 | 22 | 0.496 | 20 | 2 |
| University of Pisa | 85.7 | 65.414 | 6 | 0.764 | 5 | 1 |
| University of Camerino | 85.7 | 56.567 | 11 | 0.660 | 10 | 1 |
| University of Eastern Piedmont "A. Avogadro" | 89.9 | 50.430 | 16 | 0.561 | 15 | 1 |
| University of Siena | 84.9 | 42.211 | 20 | 0.497 | 19 | 1 |
| University of Parma | 85.9 | 39.465 | 25 | 0.459 | 24 | 1 |
| University of Trieste | 83.9 | 34.985 | 28 | 0.417 | 27 | 1 |
| University of Pavia | 84.6 | 29.109 | 36 | 0.344 | 35 | 1 |
| University of Calabria | 79.3 | 21.187 | 40 | 0.267 | 39 | 1 |
| University of Foggia | 68.0 | 15.594 | 42 | 0.229 | 41 | 1 |
| University of Naples "Parthenope" | 57.1 | 11.318 | 44 | 0.198 | 43 | 1 |
| Internation. School for Advanced Studies of Trieste | 124.5 | 265.174 | 1 | 2.130 | 1 | 0 |
| University of Messina | 87.9 | 162.387 | 2 | 1.847 | 2 | 0 |
| University of Ferrara | 89.3 | 86.471 | 3 | 0.969 | 3 | 0 |
| University of Naples "Federico II" | 84.0 | 80.117 | 4 | 0.954 | 4 | 0 |
| University of Cagliari | 85.0 | 53.967 | 12 | 0.635 | 12 | 0 |
| University of Brescia | 89.2 | 45.313 | 18 | 0.508 | 18 | 0 |
| University of Rome "La Sapienza" | 86.8 | 39.964 | 23 | 0.461 | 23 | 0 |
| University of Modena and Reggio Emilia | 86.2 | 37.264 | 26 | 0.432 | 26 | 0 |
| University of Milan "Bicocca" | 84.6 | 32.487 | 31 | 0.384 | 31 | 0 |
| Polytechnic University of Ancona | 84.6 | 30.077 | 33 | 0.356 | 33 | 0 |
| University of Molise-Campobasso | 86.1 | 28.493 | 37 | 0.331 | 37 | 0 |
| University of L'Aquila | 84.1 | 6.165 | 45 | 0.073 | 45 | 0 |
| University of Milan | 90.4 | 64.213 | 7 | 0.710 | 8 | -1 |
| University of Udine | 87.7 | 60.004 | 8 | 0.684 | 9 | -1 |
| University of Florence | 91.2 | 53.449 | 13 | 0.586 | 14 | -1 |
| University of Chieti "Gabriele D'Annunzio" | 92.2 | 51.682 | 15 | 0.560 | 16 | -1 |
| University of Genova | 90.0 | 41.539 | 21 | 0.461 | 22 | -1 |
| University of Bologna | 88.5 | 39.654 | 24 | 0.448 | 25 | -1 |
| University of Padua | 90.3 | 36.614 | 27 | 0.405 | 28 | -1 |
| University of Verona | 86.4 | 33.612 | 29 | 0.389 | 30 | -1 |
| University of Sassari | 89.8 | 29.819 | 35 | 0.332 | 36 | -1 |
| University of Rome "Tor Vergata" | 91.7 | 23.846 | 39 | 0.260 | 40 | -1 |
| University of Benevento "Sannio" | 83.1 | 17.976 | 41 | 0.216 | 42 | -1 |
| University of Palermo | 83.7 | 15.323 | 43 | 0.183 | 44 | -1 |
| University of Catanzaro "Magna Grecia" | 86.6 | 32.591 | 30 | 0.376 | 32 | -2 |
| University of Salerno | 92.9 | 45.534 | 17 | 0.490 | 21 | -4 |
| University of Catania | 93.6 | 29.964 | 34 | 0.320 | 38 | -4 |
| University of Milan "Vita-Salute San Raffaele" | 124.5 | 75.294 | 5 | 0.605 | 13 | -8 |

*Table 5: Comparison between productivity per labor unit (LP) and per unit of cost (CP) for Italian universities, for the Pharmacology SDS, 2004-2006 data.*